\documentclass[twocolumn,a4paper]{article} %<--- For `pdflatex`' user

\usepackage{nolta2025}
\usepackage{amsmath} %<--- Please uncomment this command if you use amsmath package.
\usepackage{txfonts}
\usepackage{color}

\begin{document}

\title{Slow and Fast Neurons Cooperate in Contextual Working Memory through Timescale Diversity}

\author{
  Tomoki Kurikawa${}^\dag$}

\address{
\dag Department of Complex and Intelligent systems, Future University Hakodate \\
116-2, Kameda-Nakanocho, Hakodate, Hokkaido 041-0803, Japan, \\
\\[5pt]
Email: \email{kurikawa@fun.ac.jp}
}

\maketitle

%%% Notice for ORCiD:
% Do not destroy the format, e.g., do not remove/replace the colons, spaces, or commas.
% Though do not hesitate to add another record following the form.
\orcid{
First Author: 0000-0003-1475-2812
}

\abstract
Neural systems process information across a broad range of intrinsic timescales, both within and across cortical areas. 
While such diversity is a hallmark of biological networks, its computational role in nonlinear information processing remains elusive.
In this study, we examine how heterogeneity in intrinsic neural timescales within the frontal cortex—a region central to cognitive control—enhances performance in a context-dependent working memory task.
We develop a recurrent neural network (RNN) composed of units with distinct time constants to model a delayed match-to-sample task with contextual cues.
This task demands nonlinear integration of temporally dispersed inputs and flexible behavioral adaptation. 
Our analysis shows that task performance is optimized when fast and slow timescales are appropriately balanced. 
Intriguingly, slow neurons, despite weaker encoding of task-relevant inputs, play a causal role in sustaining memory and improving performance. 
In contrast, fast neurons exhibit strong but transient encoding of input signals.
These results highlight a division of computational roles among neurons with different timescales: slow dynamics support stable internal states, while fast dynamics enable rapid signal encoding. Our findings provide a mechanistic account of how temporal heterogeneity contributes to nonlinear information processing in neural circuits, shedding light on the dynamic architecture underlying cognitive flexibility.
\endabstract

\section{Introduction}

The neural system is a complex network composed of components that operate across a broad range of timescales. Fast neural dynamics are essential for executing cognitive functions, whereas slower synaptic processes—such as synaptic plasticity—govern long-term connectivity and learning. Recent experimental studies have revealed that even in neural activity, a wide distribution of intrinsic timescales is commonly observed, both across different cortical areas\cite{Murray2014,Runyan2017} and within individual regions\cite{Cavanagh2016,Cavanagh2020,Perez-Nieves2021}.

Intrinsic timescales have been related to perceptual and working memory processes\cite{Wasmuht2018,Cavanagh2016}, yet the precise roles of neurons with distinct timescales in ongoing information processing remain poorly understood. Some studies suggest that the hierarchy of intrinsic timescales observed during resting states is maintained during task engagement, and that this temporal diversity may enhance learning\cite{Perez-Nieves2021}. However, the functional contributions of neurons with varying timescales—such as their representational differences or their roles in supporting task execution—are still largely unresolved.

Complementary modeling studies have demonstrated that neural networks incorporating neurons with multiple timescales can generate compositional sequences\cite{Yamashita2008,Kurikawa2020a,Kurikawa2021d,TomokiKurikawa2025}. In these models, slow neural dynamics serve as a ``glue'' that binds fast activity patterns, enabling the formation of structured, sequential behavior. While these studies highlight the functional importance of fast and slow neurons in cognitive tasks, their respective roles are often designed \textit{a priori} and do not emerge through learning. Thus, the question of what functional roles naturally arise from neurons with diverse timescales remains open.

In the present study, we investigate the functional significance of intrinsic neural timescales in a context-dependent working memory (CWM) task that requires the sustained maintenance of contextual information throughout each trial (Fig.1B). To this end, we train an artificial recurrent neural network (ARNN) composed of neurons with two distinct time constants and analyze how temporal heterogeneity supports task performance.

\begin{figure}[ht]
    \centering
    \includegraphics[width=1\linewidth]{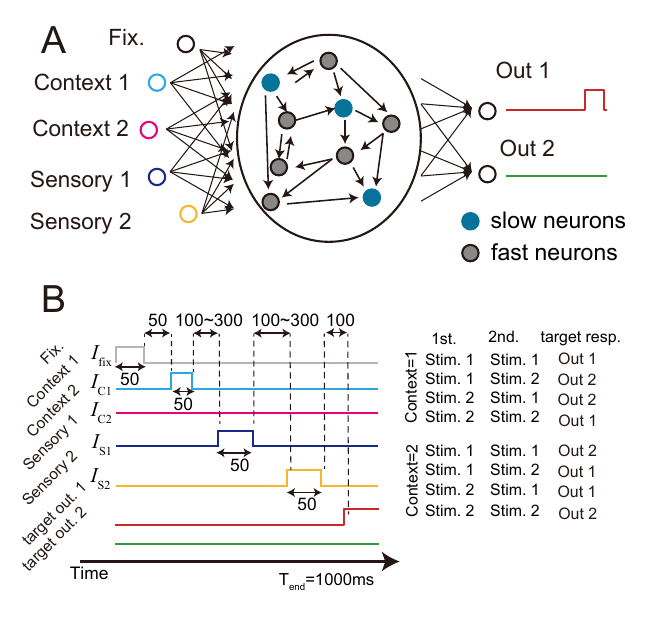}
    \caption{Image of our model and task. A: Network model image. 
    B Left: Task procedure showing a trial with context 1. B Right: A table showing the mapping between the inputs and target outputs, depending on the context.
    }
\end{figure}

\section{Methods}
\subsection{Network Model}
The recurrent neural network model consists of $N_{\mathrm{in}} = 5$, $N = 200$, and $N_{\mathrm{out}} = 2$ neurons in the input, hidden (recurrent network), and output layers, respectively. The hidden layer comprises $N$ rate-based neurons whose dynamics are governed by the following equation:

\begin{equation}
\tau_i \frac{dx_i}{dt} = \tanh\left( \sum_{j \neq i} J_{ij} x_j + \sum_j W^{\mathrm{in}}_{ij} I_j + b_i \right) - x_i,
\label{eq:neural_dyn}
\end{equation}

where $x_i$ denotes the activity of the $i$-th neuron, $\tau_i$ its intrinsic timescale, and $J$ the recurrent connectivity matrix in the hidden layer. The input vector $\mathbf{I}$ is projected to the hidden layer via the input weight matrix $W^{\mathrm{in}}$, and $b_i$ is a neuron-specific bias term. Details of the input signals $\mathbf{I}$ are provided in the next section.

The output layer comprises $N_{\mathrm{out}}$ neurons that receive input from the hidden layer through a fully connected output weight matrix $W^{\mathrm{out}}$. The output activity is given by:

\begin{equation}
\mathbf{x}_{\mathrm{out}}(t) = W^{\mathrm{out}} \mathbf{x}(t) + \mathbf{b}_{\mathrm{out}},
\end{equation}

where $\mathbf{x}(t)$ is the vector of hidden neuron activities and $\mathbf{b}_{\mathrm{out}}$ is an output bias vector.

The network is trained to perform the context-dependent working memory (CWM) task using backpropagation through time (BPTT), with trainable parameters $J, W^{\mathrm{in}}, W^{\mathrm{out}}, \mathbf{b}, \mathbf{b}_{\mathrm{out}}$. 
Equation~\eqref{eq:neural_dyn} is discretized for training, and the Adam optimizer is employed.

In this study, each hidden neuron is assigned one of two fixed intrinsic timescales, representing \textit{fast} and \textit{slow} dynamics. Fast neurons constitute 80\% of the hidden layer, and slow neurons the remaining 20\%. 
These timescale parameters are not trainable and are fixed throughout training, with $\tau_{\mathrm{fast}} = 1$ and $\tau_{\mathrm{slow}} = 10$, unless otherwise noted in analyses examining the dependence on $\tau_{\mathrm{slow}}$.

\begin{figure}
    \centering
    \includegraphics[width=1\linewidth]{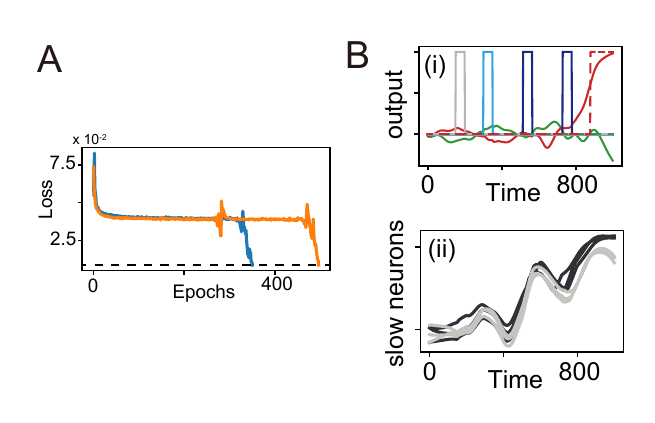}
    \caption{A: Two representative histories of the learning with the decrease in the loss function (MSE between the target output and actual output).
    The dotted line shows the threshold to terminate the training.
    B: Representative dynamics of the model after training. 
    (i): The output behavior. Red and green lines indicate the output 1 and 2 neurons, while the gray, light blue, and dark blue lines show the fixation neuron, the context 1 neuron, and the sensory 1 neuron, respectively. 
    The dotted lines show the target outputs.
    (ii): The representative neural activity of a slow neuron. Black and gray lines indicate the different conditions. Different lines represent the different trials.}
\end{figure}

\subsection{Context-dependent Working memory task (CWM)}

The input consists of five types of signals, each represented by a dedicated input neuron (see Fig. 1A). 
The fixation, context, and two sensory cues (1st and 2nd sensory inputs) are activated sequentially. 
Context is randomly chosen between contexts 1 and 2; sensory cues are also randomly selected from two options.
In context 1 trials, the network is required to produce output 1 when the first and second sensory inputs are identical, whereas in context 2 trials, it must produce output 2 under the same condition, as shown in Fig. 1B.

The durations of the activations of neurons and their intervals are shown in Fig. 1B.
When the input neuron is activated, its activity (for instance $I_\text{fix}$ for the fix, neuron) is set at 1.
The input to the recurrent network, $I$ in Eq. \ref{eq:neural_dyn}, is composed of activities of all input neurons 
$I=(I_\text{fix}, I_\text{C1}, I_\text{C2}, I_\text{S1}, I_\text{S2})^T$.

\begin{figure}
    \centering
    \includegraphics[width=0.8\linewidth]{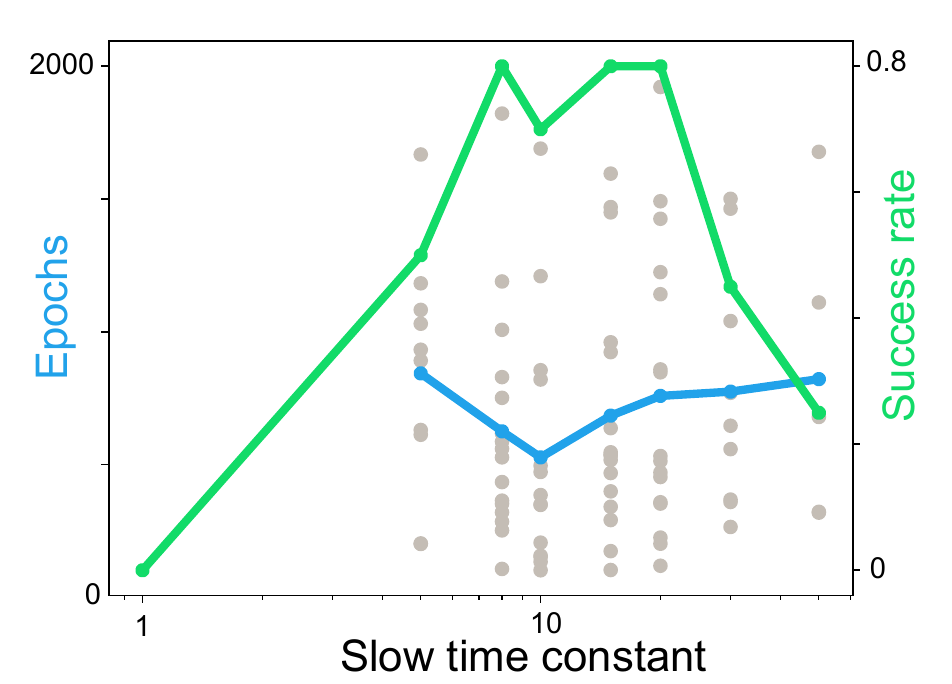}
    \caption{Training performance depending on the slow timescale $\tau_{slow}$. The blue line shows the epochs to complete training, while the green line represents the success rate across 20 realizations. Each point indicates each of the realizations of networks. For $\tau_{slow}=1$, there is no success network so that the success rate is N/A. }
\end{figure}

\section{Results}

\subsection{Impact of balancing fast and slow neural timescales}

We first examined how the balance between fast and slow neural timescales affects training performance. Training continued until the loss function reached 0.01 or until a maximum of 2000 epochs was reached (Fig. 2). Fixing $\tau_{\text{fast}} = 1$ and varying $\tau_{\text{slow}}$, we measured the training success rate and the number of epochs required for successful completion (Fig.3). Performance was optimal around $\tau_{\text{slow}} \approx 10$, as measured by both training speed and success rate. These timescale values correspond to membrane time constants. Notably, biological studies in monkeys and humans report membrane constants ranging from 10 ms to 70 ms\cite{Perez-Nieves2021}, suggesting that biological systems may be tuned for optimal cognitive performance.

\subsection{Fast neurons encode signals more effectively}

Next, we analyzed how neurons with different intrinsic timescales encode various task-related signals (context, sensory inputs, and decisions; Fig.~4A). Encoding strength, $D$, for a specific signal $A$ was quantified as
\[
D = \max_t \left| \langle x_i(t) \rangle_{A=1} - \langle x_i(t) \rangle_{A=2} \right|,
\]
where $\langle \cdot \rangle$ denotes the trial-average under the specified condition. 
For instance, when A is the context signal, $A=1$ indicates the context 1 neuron is activated.
We calculate $D$ averaged over the fast and slow neurons for the context signal and find that the fast neurons encode the context signal more strongly than the slow neurons.
Fast neurons consistently exhibited stronger encoding (mean $D \approx 0.8$ for context signals) compared to slow neurons (mean $D \approx 0.5$), confirmed across multiple networks (paired $t$-test, $p = 0.00028$).

Further, for the other signals (1st, 2nd signals, and the decisions), the fast neurons encode them strongly than the slow neurons(p=0.0035,
0.00069, 0.000023 for the 1st, 2nd signals, and the decisions, respectively).
These results demonstrate that the activities of the fast neurons change fast and, consequently, they are more influenced by application of the external signals.

\subsection{Critical contribution of slow neurons despite weaker encoding}

Given the stronger encoding observed in fast neurons, we investigated whether slow neurons are also essential for task performance. 
We selectively inactivated 10 randomly chosen neurons from either the fast or slow groups by clamping their activities to zero throughout the inference process (namely, recall process after training without change in the parameters).
For this process, we measure the mean square error between the target behaviors and the actual behaviors generated by the network with the inactivation and plot it in Fig. 4B.
The inactivation of slow neurons led to a larger increase in MSE than that of fast neurons, indicating a greater contribution of slow neurons to task performance.

The larger increase in MSE due to the inactivation of slow neurons holds for all networks.
This result shows that the slow neurons are more critical for task performance, despite their smaller encoding scale.

\subsection{Measured time scales of the neural activities}
Finally, to understand this discrepancy, we measured the time scales of the neural activity directly, as shown in Figure 4C.
Here, we computed the autocorrelation function of the activity for each neuron and then identify the time scale of the neural activity by fitting the autocorrelation curve with the exponential curve.
Thus, the timescales of the neurons are measured independent of the time constants $\tau_{fast}$ and $\tau_{slow}$.
We measured the timescale for $\tau_{fast}=1$ and $\tau_{slow}=10$ and plot it in Fig. 4C.
We found that slow neurons exhibited longer empirical timescales (mean $\approx 38$) than fast neurons (mean $\approx 23$). Notably, the ratio of empirical timescales (approximately 1.5) was much smaller than the ratio of their predefined time constants (10).

Moreover, empirical timescales of slow neurons varied only moderately with changes in $\tau_{\text{slow}}$ (e.g., for $\tau_{\text{slow}} = 5, 10, 20$, the measured timescales were 33, 59, and 66, respectively). These findings suggest that the temporal structure of the task plays a major role in shaping the actual dynamics of slow neurons, beyond what is imposed by their intrinsic time constants.

\begin{figure*}
    \centering
    \includegraphics[width=0.9\linewidth]{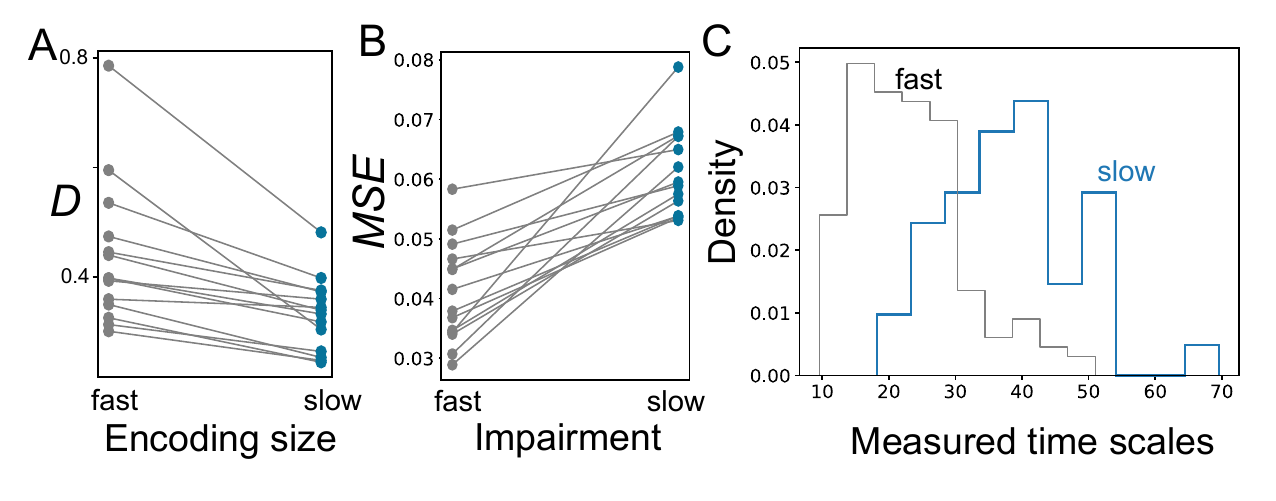}
    \caption{A: encoding size for the context signal. $D$ of the fast and slow neurons are averaged over the fast and slow neurons, respectively.
    B: The mean square error (MSE) by the inactivation of the fast and slow neurons. MSE without the inactivation is less than $0.01$. The lines in A and B indicate the different networks.
    C: The measured timescales of neurons. For each neuron, the timescale of its activity is measured. See the main text.}
\end{figure*}

\section{Conclusion}
This study demonstrates that an optimal balance between fast and slow neural timescales is essential for effective performance in context-dependent working memory tasks. 
Although fast neurons exhibit stronger encoding of task-relevant signals, our analyses reveal that slow neurons play a more critical causal role in maintaining stable memory and ensuring successful task execution.
The distinct contributions of fast and slow neurons point to a functional division of labor: fast neurons encode task information transiently and precisely, while slow neurons sustain internal states necessary for ongoing computations. 
This supports the idea that task-relevant signals are first encoded by fast neurons and then relayed to slow neurons, enabling stable maintenance over time.

Overall, our results suggest that temporal heterogeneity within neural populations enhances cognitive performance by integrating rapid encoding and persistent memory functions. 
This work offers a mechanistic explanation for how diverse neural timescales can support complex cognitive computations in biologically inspired systems.

\section*{Acknowledgments}
The present work is supported by JSPS KAKENHI (25H02622), and Special Research Expenses in Future University Hakodate

\bibliographystyle{ieeetr}
\bibliography{nolta2025}

\begin{thebibliography}{10}

\bibitem{Murray2014}
J.~D. Murray, A.~Bernacchia, D.~J. Freedman, R.~Romo, J.~D. Wallis, X.~Cai, C.~Padoa-Schioppa, T.~Pasternak, H.~Seo, D.~Lee, and X.-J. Wang, ``{A hierarchy of intrinsic timescales across primate cortex.},'' {\em Nature neuroscience}, vol.~17, pp.~1661--1663, nov 2014.

\bibitem{Runyan2017}
C.~A. Runyan, E.~Piasini, S.~Panzeri, and C.~D. Harvey, ``{Distinct timescales of population coding across cortex},'' {\em Nature}, vol.~548, pp.~92--96, aug 2017.

\bibitem{Cavanagh2016}
S.~E. Cavanagh, J.~D. Wallis, S.~W. Kennerley, and L.~T. Hunt, ``{Autocorrelation structure at rest predicts value correlates of single neurons during reward-guided choice},'' {\em eLife}, vol.~5, no.~OCTOBER2016, pp.~1--17, 2016.

\bibitem{Cavanagh2020}
S.~E. Cavanagh, L.~T. Hunt, and S.~W. Kennerley, ``{A Diversity of Intrinsic Timescales Underlie Neural Computations},'' {\em Frontiers in Neural Circuits}, vol.~14, no.~December, pp.~1--18, 2020.

\bibitem{Perez-Nieves2021}
N.~Perez-Nieves, V.~C. Leung, P.~L. Dragotti, and D.~F. Goodman, ``{Neural heterogeneity promotes robust learning},'' {\em Nature Communications}, vol.~12, no.~1, pp.~1--9, 2021.

\bibitem{Wasmuht2018}
D.~F. Wasmuht, E.~Spaak, T.~J. Buschman, E.~K. Miller, and M.~G. Stokes, ``{Intrinsic neuronal dynamics predict distinct functional roles during working memory},'' {\em Nature Communications}, vol.~9, p.~3499, dec 2018.

\bibitem{Yamashita2008}
Y.~Yamashita and J.~Tani, ``{Emergence of functional hierarchy in a multiple timescale neural network model: a humanoid robot experiment.},'' {\em PLoS computational biology}, vol.~4, p.~e1000220, nov 2008.

\bibitem{Kurikawa2020a}
T.~Kurikawa and K.~Kaneko, ``{Multiple-Timescale Neural Networks: Generation of History-Dependent Sequences and Inference Through Autonomous Bifurcations.},'' {\em Frontiers in computational neuroscience}, vol.~15, p.~743537, jun 2021.

\bibitem{Kurikawa2021d}
T.~Kurikawa, ``{Intermediate Sensitivity of Neural Activities Induces the Optimal Learning Speed in a Multiple-Timescale Neural Activity Model},'' in {\em Neural Information Processing. ICONIP 2021} (A.~{Mantoro, T., Lee, M., Ayu, M.A., Wong, K.W., Hidayanto}, ed.), pp.~64--72, Springer, Cham, 2021.

\bibitem{TomokiKurikawa2025}
K.~K. {Tomoki Kurikawa}, ``{Stability Control of Metastable States as a Unified Mechanism for Flexible Temporal Modulation in Cognitive Processing},'' {\em arXiv preprint}, vol.~2504.09080, 2025.

\end{thebibliography}

\end{document}